\documentclass{pasj01}
\Received{$\langle$reception date$\rangle$}
\Accepted{$\langle$acception date$\rangle$}
\Published{$\langle$publication date$\rangle$}

\begin{document}

\title{First results on the cluster galaxy population from the Subaru Hyper Suprime-Cam survey. I. The role of group or cluster environment in star formation quenching from $z$ = 0.2 to 1.1}
\author{Hung-Yu Jian$^{1*}$, Lihwai Lin$^{1}$, Masamune Oguri$^{2,3,4}$, Atsushi Nishizawa$^{5}$ , Masahiro Takada$^{4}$, Surhud More$^{4}$, Yusei Koyama$^{6,7}$, Masayuki Tanaka$^{8}$, Yutaka \textsc{Komiyama}\altaffilmark{8,9}}%
\altaffiltext{1}{Institute of Astronomy \& Astrophysics, Academia Sinica, 106, Taipei, Taiwan, R.O.C.}
\altaffiltext{2}{Research Center for the Early Universe, University of Tokyo, Tokyo 113-0033, Japan}
\altaffiltext{3}{Department of Physics, University of Tokyo, Tokyo 113-0033, Japan}
\altaffiltext{4}{Kavli Institute for the Physics and Mathematics of the Universe (Kavli IPMU, WPI), Universityof Tokyo, Chiba 277-8582, Japan}
\altaffiltext{5}{Institute for Advanced Study, Nagoya University, Furocho Chikusaku Nagoya 4648602, Aichi,Japan}
\altaffiltext{6}{Department of Astronomical Science, SOKENDAI, Osawa, Mitaka, Tokyo 181-8588, Japan}
\altaffiltext{7}{Subaru Telescope, National Astronomical Observatory of Japan, 650 North A’ohoku Place, Hilo, HI 96720, USA}
\altaffiltext{8}{National Astronomical Observatory of Japan, Osawa 2-21-1, Mitaka, Tokyo 181-8588, Japan}
\altaffiltext{9}{The Graduate University for Advanced Studies, 2-21-1 Osawa, Mitaka, Tokyo 181-8588, Japan}

\email{hyjian@asiaa.sinica.edu.tw}


\maketitle

\begin{abstract}
We utilize the HSC CAMIRA cluster catalog and the photo-$z$ galaxy catalog constructed in the HSC wide field (S16A), covering $\sim$ 174 deg$^{2}$, to study the star formation activity of galaxies in different environments over 0.2 $<$ $z$ $<$ 1.1. We probe galaxies down to $i \sim$ 26, corresponding to a stellar mass limit of  log$_{10}$(M$_*$/M$_{\odot}$) $\sim$ 8.2 and $\sim$ 8.6 for star-forming and quiescent populations, respectively, at $z$ $\sim$ 0.2. The existence of the red sequence for low stellar mass galaxies in clusters suggests that the environmental quenching persists to halt the star formation in the low-mass regime. In addition, star-forming galaxies in groups or clusters are systematically biased toward lower values of specific star formation rate by 0.1 -- 0.3 dex with respect to those in the field and the offsets shows no strong redshift evolution over our redshift range, implying a universal slow quenching mechanism acting in the dense environments since $z$ $\sim$ 1.1. Moreover, the environmental quenching dominates the mass quenching in low mass galaxies, and the quenching dominance reverses in high mass ones. The transition mass is greater in clusters than in groups, indicating that the environmental quenching is more effective for massive galaxies in clusters compared to groups.

\end{abstract}

\section{Introduction}

It is well established that clusters at $z < 1$ are dominated by galaxies with redder colors, older stellar populations, early type morphologies, and little star formation, as opposed to the field environments. Many scenarios related to cluster environments have been proposed to explain how the star formation is ceased in clusters, including processes, such as ram-pressure stripping \citep{gun72,dre83}, and galaxy-galaxy interaction \citep{mih94} that quench star formation over a short timescale, and `strangulation', referring to the removal of warm and hot gas \citep{lar80,bal00}, which slowly reduces the cold gas supply. Studying the quenching timescale for cluster galaxies hence provides a powerful way to constrain the physical mechanisms responsible for the star formation quenching.

One of the tools to constrain the quenching timescale is the comparison of the properties of star-forming galaxies between different environments. It is found that star-forming galaxies (hereafter, SF population) form a tight sequence on the SFR and stellar mass plane, the so-called `main sequence' \citep{bri04,noe07,elb07,dad07,pan09,mag10,lin12,whi12,hei14}. It is expected that slow quenching would lead to an overall reduction of the specific star formation rates (sSFRs) of the SF population whereas a fast quenching mainly changes the fraction of quenched population without altering the sSFR of the remaining SF galaxies.
Previous studies have suggested that the difference in the star formation activities between the field and the cluster environments is primarily driven by the relative red fraction, instead of the properties of the SF population \citep{muz12,koy13,lin14,wag17,jian17}. Using Pan-STARRS1 clusters \citep{lin14,jian17}, it is found that the specific star formation rates (sSFRs) of SF galaxies in the clusters are only moderately lower than those the field ($<$ 0.2-0.3 dex) and the difference becomes insignificant on groups scales. However, the uncertainty remains large for high mass galaxies because of the small sample sizes used in these studies, and it is also uncertain that the environmental quenching can extend to low mass galaxies.   

In this work, we probe the properties of cluster galaxies using a large sample drawn from the HSC Subaru Strategic Program (hereafter the HSC survey). The HSC survey \citep{aih17} consists of three survey layers, `Wide', `Deep', and `UltraDeep' components. The cluster sample used in this work comes from the HSC CAMIRA (Cluster finding Algorithm based on Multi-band Identification of Red-sequence gAlaxies) catalog \citep{ogu17} constructed in the HSC Wide field, reaching $i \sim 26$ at 5$\sigma$ over 174 deg$^{2}$. It contains 4972 clusters with richness larger than 10 in the redshift range of $0.2 < z < 1.1$. This allows us not only to extend the analysis to redshifts greater than 0.8 and to fainter galaxy populations by 2 magnitudes in $i$, but also to increase the sample size by a factor of 17 compared to our previous works \citep{lin14,jian17}. There are two other companion papers in this special issue to address the environmental effects but using different catalogs and methods. \citet{koy17} make use of the NB emitter catalogs constructed for HSC-SSP Deep and UltraDeep fields to search for ``red emitter'' along the large-scale structures at 0.2 $<$ $z$ $<$ 1.7 and identify their environments. \citet{nis17} utilize the CAMIRA catalog to study the evolution of cluster profile for red and blue populations out to $z$ $<$ 1.1 and reveal the color-magnitude relation of red-sequence galaxies at the faint end $z$ $<$ 24.  

Our paper is formatted as follows. In Section 2, we briefly describe the data, the sample selection, and the analysis method. In Section 3, we present the main results, discussing the main sequence properties, and the redshift and mass dependence of field, group, and cluster galaxy properties. In Section 4, our conclusion and discussion are presented. Throughout this paper we adopt the following cosmology: Throughout this paper we adopt the following cosmology: \textit{H}$_0$ = 100$h$~km s$^{-1}$ Mpc$^{-1}$, $\Omega_{\rm m} = 0.3$ and $\Omega_{\Lambda } = 0.7$. We adopt the Hubble constant $h$ = 0.7 when calculating rest-frame magnitudes. All magnitudes are given in the AB system.

\section{Data, Sample Selection, and Method}

\subsection{HSC Galaxy Sample}
Hyper Suprime-Cam (HSC) Survey is conducted as part of a 300-night Strategic Survey Program (SSP) over 5 years starting from March of 2014, aiming to explore the nature of dark matter and dark energy as well as the evolution of galaxies. The first public data of HSC-SSP are released recently \citep{aih17}. The Survey utilizes the wide field of view of 1.77 square degrees of Hyper Suprime-Cam to collect broad-band images in \textit{grizy} bands as well as to study emission line objects at high redshifts through a number of narrow-band filters \citep{aih17}.  HSC Survey consists of a three-layered imaging, including Wide, Deep, and UltraDeep. The aimed coverage for the Wide survey has 1,400 square degrees of the sky in all the five broad-band filters located around the equator and two large stripes around the spring and autumn equator with an additional stripe around the Hectomap region. The Deep survey is carried out in four separate fields; XMM-LSS, Extended-COSMOS (ECOSMOS), ELAIS-N1 and DEEP2-F3, and the UltraDeep has two fields, COSMOS and SXDS. The detail survey description can be found in \citet{aih17}.

In this study, we make use of the S16A photometric redshift catalog based on the S16A internal data release of the HSC Survey released in 2016 August. The HSC Wide S16A dataset contains imaging data taken between 2014 March and 2016 April in all five broad-bands at full depth and covers $\sim$ 174 deg$^2$. The HSC data is processed by the HSC Pipeline, or hscPipe (Bosch et al. 2017, in prep.), which is based on the Large Synoptic Survey Telescope pipeline \citep{ive08,axe10,jur15}. In addition, the HSC astrometry and photometry are calibrated against the Pan-STARRS1 3$\pi$ catalog \citep{ton12,sch12,mag13}. For more detail descriptions, readers are referred to \citet{aih17}.

\begin{figure*}
\includegraphics[scale=1.4]{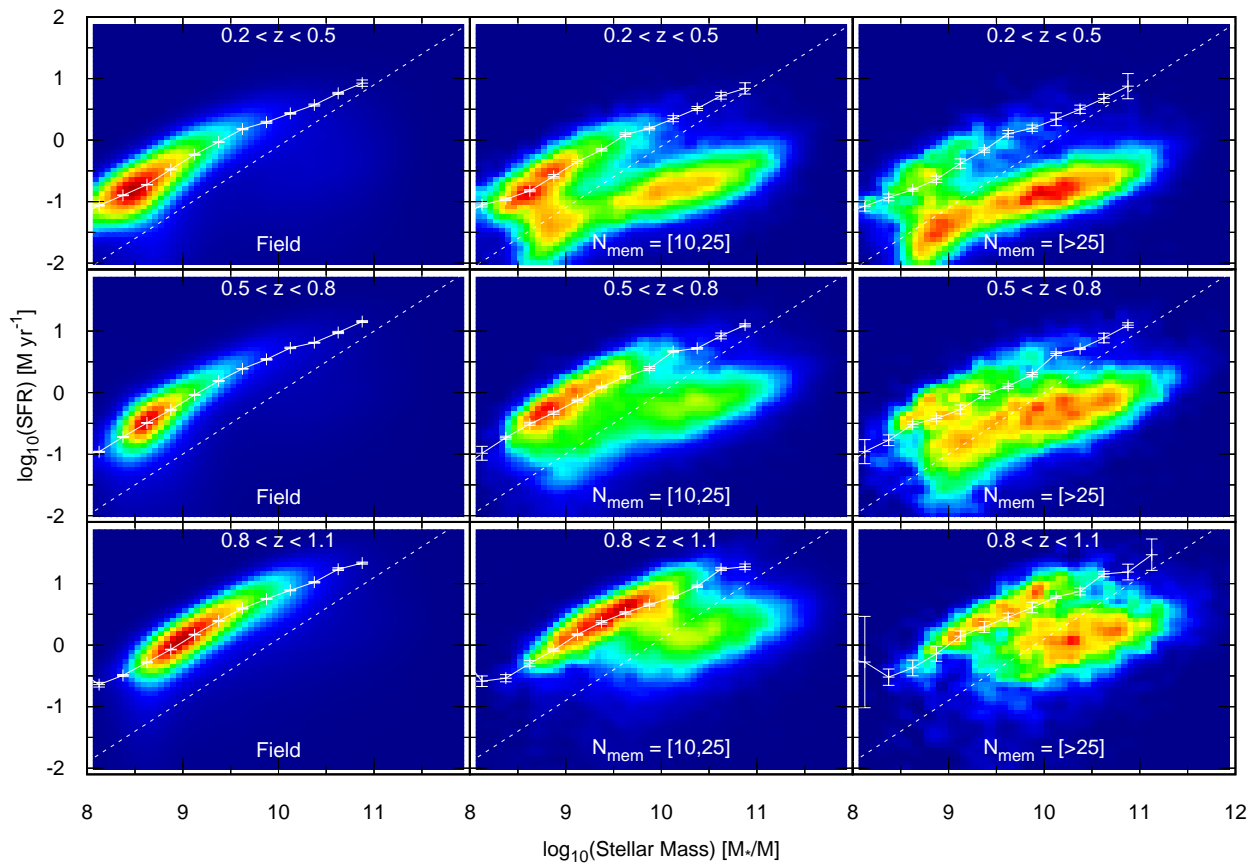}
\caption{The images show the color-coded number densities of stacked galaxies for the SFR-M$_*$ relation in three redshift ranges from [0.2,0.5], [0.5, 08] to [0.8, 1.1] (from the top to the bottom)  and in three different environments, including the field (left), groups (middle), and clusters (right), respectively. The color scale in the color-bar is proportional to the number counts in each cell normalized by the cell with the maximum count. The diagonal lines represent the boundaries to separating star-forming and quiescent populations. They are sSFR = 10.1 in $0.2 < z < 0.5$, 10. in $0.5 < z < 0.8$, and 9.9 in $0.8 < z < 1.1$, respectively. The white solid lines with error-bars, estimated using the jack-knife resampling from eight sub-samples, denote the median values of SFRs at different stellar masses. We can qualitatively learn the redshift evolution of galaxies in three distinct environments, and at the same redshift, the differences of galaxy distributions in the SFR-M$_*$ space in different environments to see the environmental impact.}
\label{f1}
\end{figure*}

\subsection{CAMIRA Groups/Clusters}
The group/cluster sample is a product produced by a cluster-finding algorithm CAMIRA (Cluster finding algorithm based on Multi-band Identification of Red sequence gAlaxies) as described in \citet{ogu14}. Based on the stellar population synthesis (SPS) model of \citet{bru03}, CAMIRA fits all photometric galaxies for an arbitrary set of bandpass filters and computes likelihoods of being red-sequence galaxies as a function of redshift. To improve the accuracy of the model prediction, additional calibration is done through using a sample of spectroscopic galaxies to derive residual colors of SPS model fitting as a function of wavelength and redshift. The detailed methodology of the CAMIRA algorithm can be found in \citet{ogu14}.

In this work, we utilize a new optically selected CAMIRA cluster catalog from the first two years of observation of the HSC Survey, i.e. the HSC Wide S16A CAMIRA catalog \citep{ogu17}, to study galaxies properties in groups and clusters. The sample contains 4972 clusters/groups with richness $N_{mem}$ $>$ 10 in redshift range of 0.1 $<$ $z$ $<$ 1.1, and is validated through comparisons with spectroscopic and X-ray data as well as mock galaxy catalogs. It is shown in \citet{ogu17} that the redshift evolution of the mass threshold is not strong. That is, the constant richness cut roughly corresponds to the constant halo mass cut. Based on the cluster redshifts and the richness $N_{mem}$, we bin the clusters into three redshift ranges,  0.2 $<$ z $<$ 0.5, 0.5 $<$ z $<$ 0.8, and 0.8 $<$ z $<$ 1.1, to study their evolution and two richness ranges, 10 $<$ $N_{mem}$ $<$25 and $N_{mem}$ $>$ 25, to explore the ``group'' and ``cluster'' environment, respectively. Here $N_{mem}$ = 10 and 25 correspond to the virial halo mass log$_{10}$(M$_{vir}$/h$^{-1}$ M$_{\odot}$) $\sim$ 13.61 $\pm$ 0.13 and 14.19 $\pm$ 0.02, based on Equation 40 in \citet{ogu14}. Be noted that clusters with richness less than 15 are not included in \citet{ogu17} since these less massive clusters (or groups) suffer more contamination. 

\begin{table}
\tbl{CAMIRA Cluster Catalog}{%
\begin{tabular}{cccc}  
\hline\noalign{\vskip3pt} 
\multicolumn{1}{c}{Redshift} & $z_{median}$ & Group & Cluster \\   
\hline\noalign{\vskip3pt}
\multicolumn{1}{c}{} & & 10 $<$ $N_{mem}$ $<$ 25 & $N_{mem}$ $>$ 25 \\  [2pt]
\multicolumn{1}{c}{} & & M$_{vir}$/M$_{\odot}$ = 10$^{13.6-14.2}$ &  10$^{> 14.2}$ \\
\hline\noalign{\vskip3pt} 
 0.2 $< z <$ 0.5  & 0.33 & 1139 & 194   \\
 0.5 $< z <$ 0.8  & 0.68 & 1506 & 153  \\
 0.8 $< z <$ 1.1  & 0.92 & 1611 & 95  \\
\hline\noalign{\vskip3pt} 
\end{tabular}}\label{table:cataloginfo}
 
\end{table}

\begin{figure}
\includegraphics[scale=0.6]{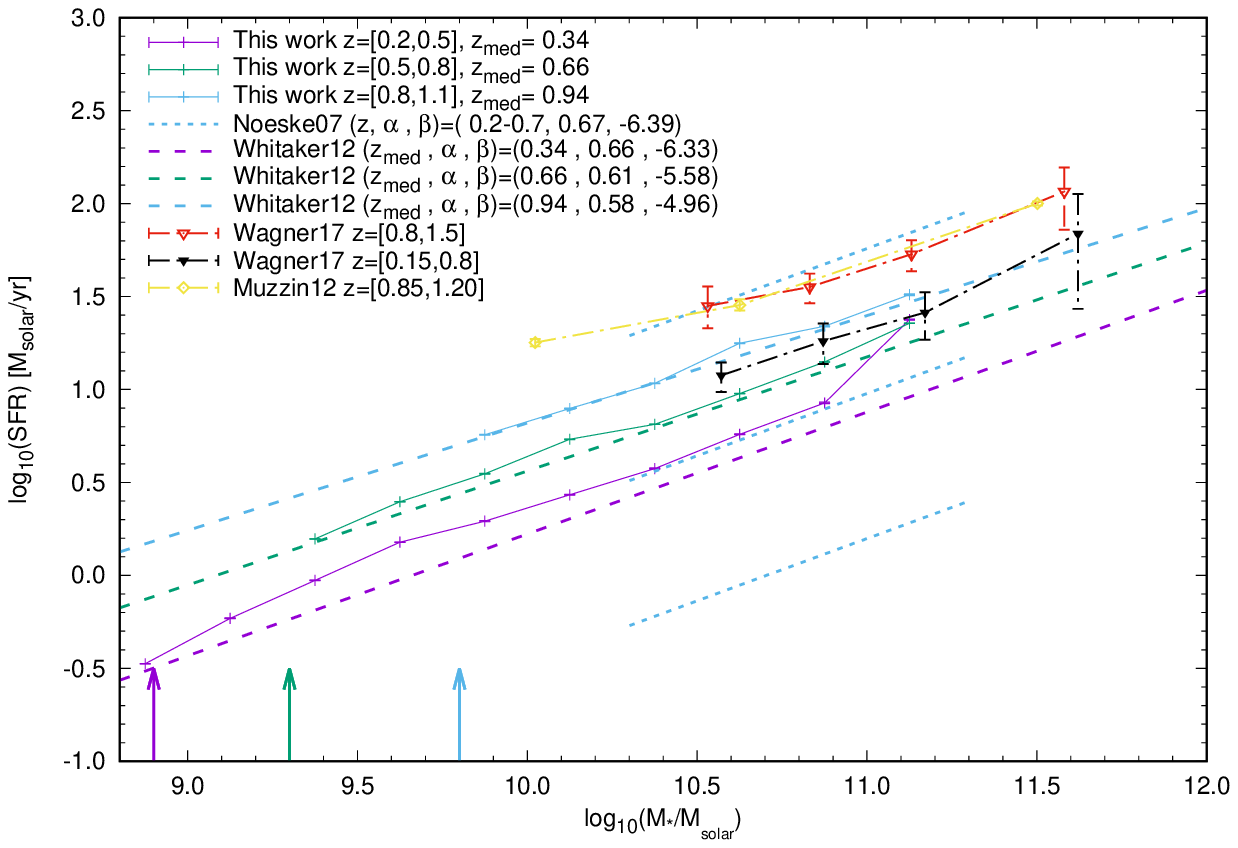}
\caption{The median SFR of the SF main sequence in the field in redshift range of 0.2 $<$ $z$ $<$ 0.5 (purple), 0.5 $<$ $z$ $<$ 0.8 (green), and 0.8 $<$ $z$ $<$ 1.1 (blue) from this work (solid lines) and from \citet{whi12} (dashed lines), respectively. The results of the SF main sequence in the field from \citet{wag17} (red and black triangles), \citet{muz12} (yellow), and \citet{noe07} with upper and lower limits (blue dotted lines) are also included for comparisons. The purple, green, and blue arrows denote the mass completeness limit for SF galaxies at $z$ = 0.5, 0.8, and 1.1. Our results, in general, are in agreement with the results from literature. }
\label{f2}
\end{figure}

\subsection{The photo-z catalog and stellar mass estimation}

Photometric redshifts and physical properties of galaxies such as stellar mass and star formation rate are inferred using a photometric redshift code \texttt{Mizuki} \citep{tan15}. It is a template fitting code and templates are generated using \citet{bru03}. When deriving stellar masses and star formation rates, the Chabrier initial mass function (IMF) is assumed \citep{cha03}. The code applies a set of Bayesian priors on physical properties of galaxies in order to keep the model parameters within realistic ranges.  These priors are a function of redshift, which effectively let the templates evolve with redshift. It also applies template error functions to reduce systematic biases in the model templates and also to assign uncertainties to the templates. The code is run using the $grizy$ CModel photometry.  The redshift and physical parameters are estimated by marginalizing over all the other parameters. The quoted uncertainties in the physical properties thus include uncertainties in photometric redshifts.  Details of the calibration of the codes and the data products are described in the photo-z release paper (Tanaka et al. in prep.). It is also noted that there is a known bias in stellar mass in \texttt{Mizuki} which is about a factor of 2 larger than the mass in COSMOS at $z$ $\sim$ 1 \citep{tan15}. 

\subsection{K-correction, Completeness Limit of Stellar Mass, and Star Formation Rate}
Our approach to computing the K-correction follows a similar method described in \citet{wil06} to relate the observed color and magnitude at redshift $z$ to the rest-frame color and magnitude based on empirical templates from \citet{kin96}. The detail description can be found in \citet{wil06} as well as in \citet{lin14}. In short, for a given redshift, we perform a polynomial fit between the K-correction term and a pair of adjacent observed color base on Kinney et al. templates, where the K-correction term is chosen to be the observed bandpass closest to the desired rest-frame quantity. Different input redshifts lead to different fitting results or polynomial formulas, and a table of fitting polynomial values can then be constructed and be applied to galaxies depending on their redshift in the range of 0 $<$ $z$ $<$ 1.45. 

The estimation of the stellar mass limit follows the method described in \citet{lin14}.  We first compute the rest-frame quantities for galaxies at a given redshift based on their 5σ limiting magnitudes in the observed HSC bands utilizing the K-correction method illustrated previously. Adopting the empirical formula obtained by \citet{lin07}, we then convert the rest-frame magnitudes and colors to the corresponding stellar mass. It is known that at a fixed rest-frame magnitudes, the redder color a galaxy, the higher mass.  By taking the reddest colors of star-forming and quiescent populations, we derive their corresponding mass limits. In this way, we find that the mass limits log$_{10}$ (M$_*$ /M$_{\odot}$) are 8.6 (8.2), 9.3 (8.9), 9.7 (9.3) and 10.3 (9.8) for red (blue) galaxies at z $\sim$ 0.2, 0.5, 0.8, and 1.1, respectively.

\begin{figure*} 
\includegraphics[scale=1.4]{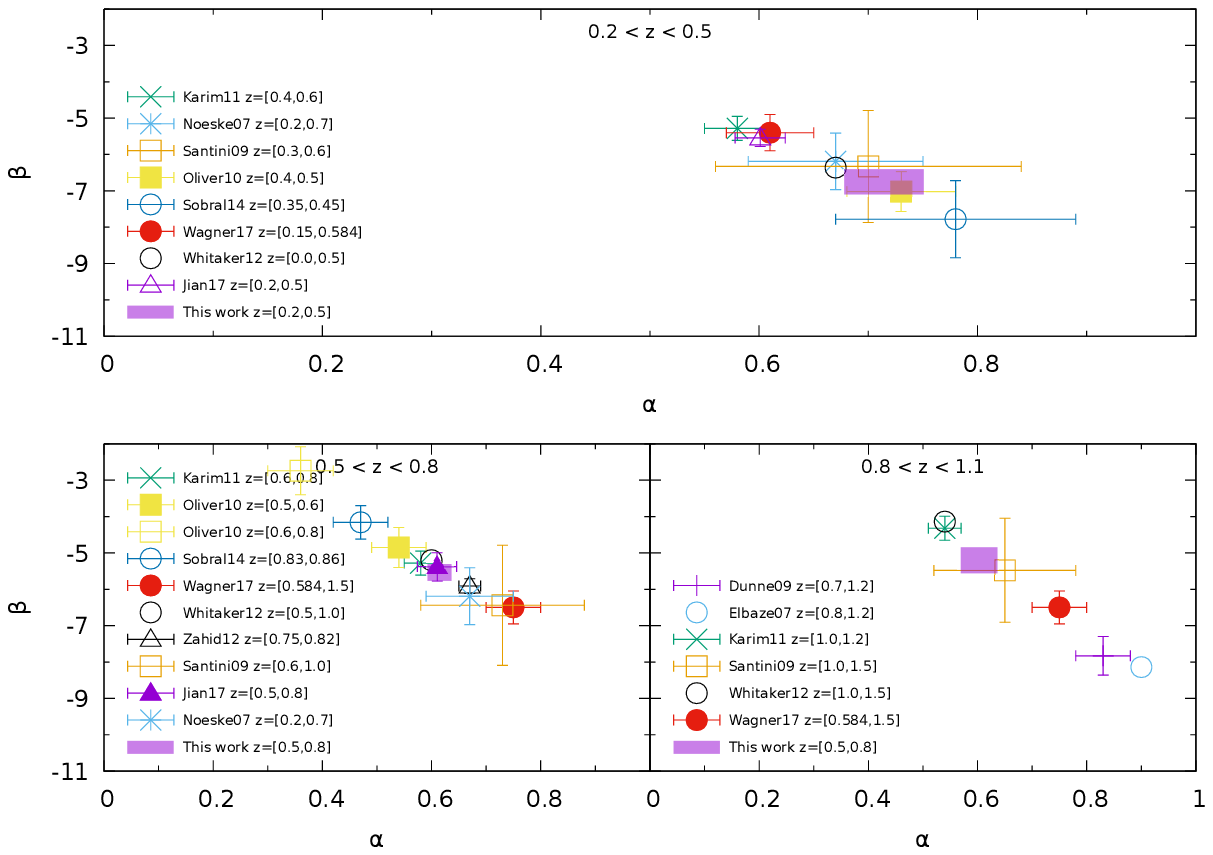}
\caption{Best-fit parameters $\alpha$ and $\beta$ in Equation~\ref{equ1} for SF galaxies are plotted in three redshift ranges. This figure is based on the work by \citet{spe14}, who compile and calibrate the best-fit parameters from literatures including \citet{elb07}, \citet{noe07}, \citet{dun09}, \citet{san09}, \citet{oli10}, \citet{kar11}, \citet{whi12}, \citet{zah12}, and \citet{sob14}. Additionally, we add the results from \citet{wag17}, \citet{jian17} and this work. The best-fit parameters from \citet{whi12} are without uncertainties and shown by open circles.}
\label{f3}
\end{figure*}

We compute the SFR following the approach described in \citet{mos12}. The SFR is parameterized as functions of rest-frame optical B magnitudes M$_B$ and (U-B) color and calibrated against SED-fit SFR  from UV/optical bands in the All-Wavelength Extended Groth Strip International Survey for both red sequence and blue cloud galaxies in the 0.7 $<$ $z$ $<$ 1.4 redshift range. The calculated SFRs are found to agree well with an L[O$_{\rm{II}}$]-M$_B$ SFR calibration commonly used in the literature by considering a correction for the measured values M$_B$ from DEEP2 galaxies to include a dimming factor of Q = 1.3 magnitudes per unit redshift \citep{mos12}. In this work, we adopt the fitting formula using the rest-frame optical M$_B$, (U-B), and a second order (U-B) color as fit parameters and the fit coefficients can be found in Table 3 in \citet{mos12}. It is reported in \citet{mos12} that the SFR uncertainties depend on colors of galaxies. Although the method gives more precise SFRs for star-forming galaxies and less accurate SFRs for quiescent ones, it uncovers the SFR of galaxies with a wide range of star formation activities. In this study, the main emphases are the comparisons of star formation rate for star-forming galaxies and the quiescent fraction in different environments. The larger uncertainty in the SFR estimate for quiescent objects has little impact on our results. 


The sSFR, defined as SFR/M$_*$, is used to separate star-forming and quiescent populations. In this work, the separating thresholds we apply are log$_{10}$(sSFR) = -10.1 yr$^{-1}$ in $0.2 < z < 0.5$, -10.0 yr$^{-1}$ in $0.5 < z < 0.8$, and -9.9 yr$^{-1}$ in $0.8 < z < 1.1$.  

\subsection{Method}
 
By construction, the CAMIRA cluster catalog only contains red cluster members. In order to probe the properties of general populations in clusters, we perform the stacking analysis and correct for the background/foreground contaminations. The details of the method are discussed and illustrated in \citet{jian17}.  In short, the contaminated galaxy properties around the cluster centers during stacking can be statistically removed by subtracting the same galaxy properties of a mean local background, selected either from random positions or from annuluses between $r_1$ and $r_2$, where $r_1$ and $r_2$ are the inner and the outer comoving radii around cluster centers, respectively. For each cluster, we project galaxies around its center onto a plane within a redshift width equal to the photo-$z$ accuracy of the galaxy sample, i.e. $|z - z_{\textrm{grp}}| \leq \sigma_{\Delta z/(1+z_s)}$, where $z$, $z_{\textrm{grp}}$, and $\sigma_{\Delta z/(1+z_s)}$ are galaxy, group redshift, and photo-$z$ accuracy, respectively. For galaxies on this projected plane, their redshift is adjusted to the cluster redshift for k-correction. Galaxies within a projection radius $r_p$ 1.5 $\textrm{Mpc}$ from the center are considered as the contaminated cluster galaxy sample. The corrected sample can be obtained by subtracting the contaminated cluster galaxy sample with an area-normalized background sample. In this work, we select comoving $r_1$ and $r_2$ to be 8.0 and 10.0 $\textrm{Mpc}$, respectively.

\begin{figure*}
\includegraphics[scale=0.6]{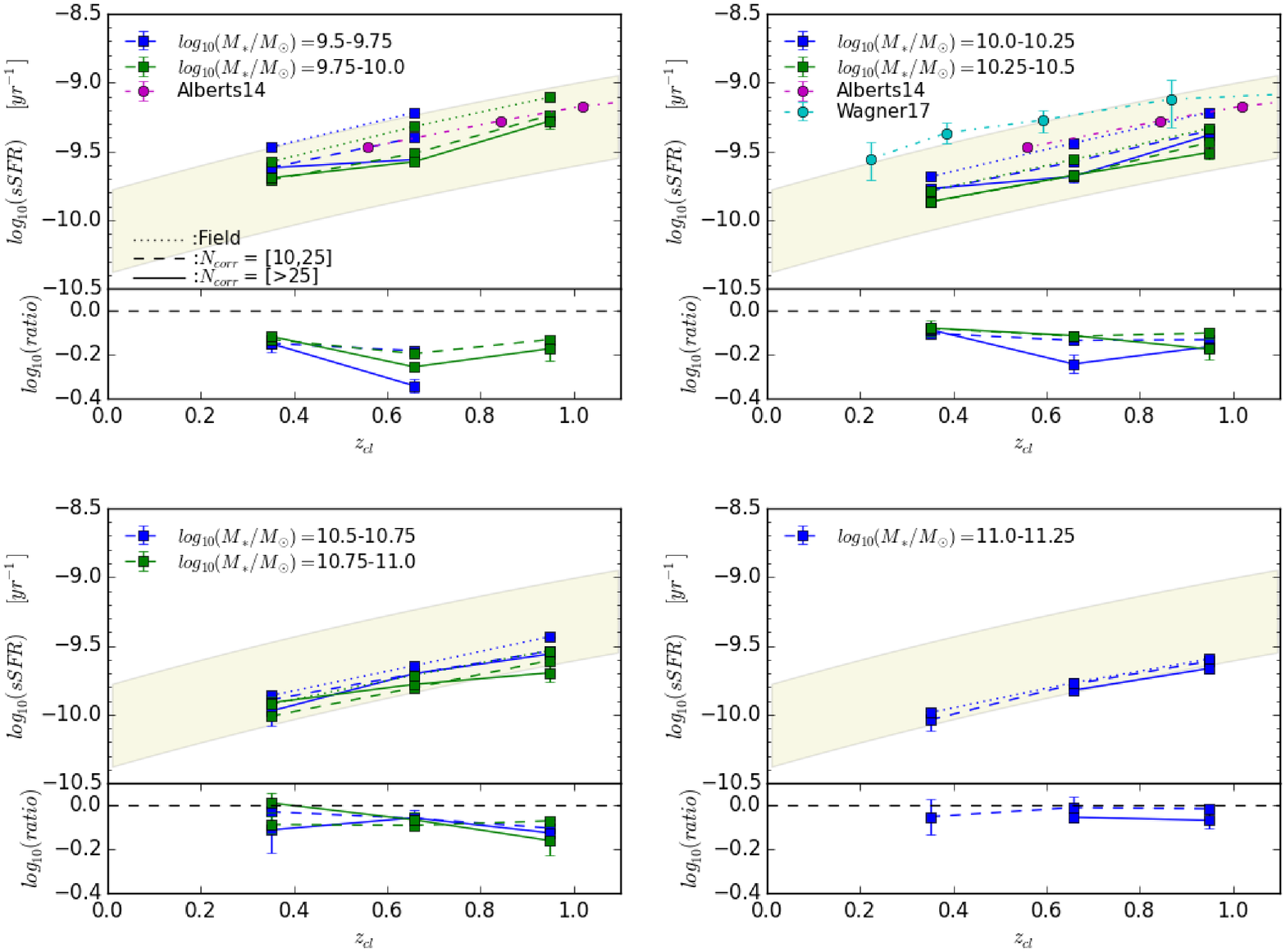}
    \caption{Top: The median sSFR of SF galaxies as a function of redshift in the field (dotted), groups (dashed), clusters (solid), respectively. Bottom: The logarithmic ratio of sSFR for group to field galaxies (dashed), and for cluster to field galaxies (solid). In addition, the red and cyan circles denote the sSFR of SF galaxies in the field from \citet{alb14} for M$_*$ $>$ 1.3 $\times$ 10$^{10}$ M$_{\odot}$ and in clusters from \citet{wag17} for M$_*$  $>$ 10$^{10.1}$ M$_{\odot}$, respectively. The yellow shade region gives the boundaries of the infrared Main Sequence as defined in \citet{elb11}. In the upper left panel, the data points in the redshift range of 0.8 $<$ $z$ $<$ 1.1 for the mass bin 9.5-9.75 are removed due to the mass incompleteness. The trend that the sSFR decreases with the decreasing redshift and the decreasing level for the same redshift range in our results are in good agreement with other works although a systematics is seen.}
\label{f4}
\end{figure*}

\section{Results}
\subsection{Main Sequence Properties}
The main sequence indicates the tight correlation between galaxy SFR and M$_{*}$ for star-forming (SF) galaxies.  In Figure~\ref{f1}, we show the  color-coded number density plots of stacked galaxies for the SFR-M$_{*}$ relation in 3 redshift bins,[0.2,0.5], [0.5,08] and [0.8,1.1] (from the top to the bottom), and in the environment of the field (left), groups (middle), and clusters (right). The diagonal dashed lines separate the star-forming or the main sequence (above) and quiescent galaxies (below) with log$_{10}$(sSFR) = -10.1 yr$^{-1}$ in $0.2 < z < 0.5$, -10.0 yr$^{-1}$ in $0.5 < z < 0.8$, and -9.9 yr$^{-1}$ in $0.8 < z < 1.1$. The colors are scaled with the number counts in each cell normalized by the cell with the maximum count. The solid white lines with error-bars give the median SFRs of SF galaxies in different redshifts and environments as a function of the stellar mass. The error-bars are estimated from eight sub-samples using the jack-knife resampling method. It is seen that the distributions of field galaxies have a distinct appearance from those of group or cluster galaxies. In the field, the presence of star-forming galaxies is so prominent that quiescent galaxies can almost be neglected for all masses. By contrast, in groups or clusters, quiescent galaxies become the dominant ones for mass roughly larger than 10$^{9.5-10}$ $M_{\odot}$. In general, our results are in good agreement with the conclusions from previous studies that the group or cluster environment has a higher red fraction of galaxies than the field environment \citep{ger07,gio12}. In addition, the evolution of field, group, and cluster galaxies can also be seen clearly. Qualitatively, the main sequence of SF field galaxies has lower sSFR as it evolves from high redshift to low redshift. The fraction of the quiescent population in groups and clusters becomes higher with the decreasing redshift, and at the same redshift, the quiescent population in clusters is more prominent than in groups, evident the hosting environmental effect. 

\begin{figure*}
\includegraphics[scale=0.6]{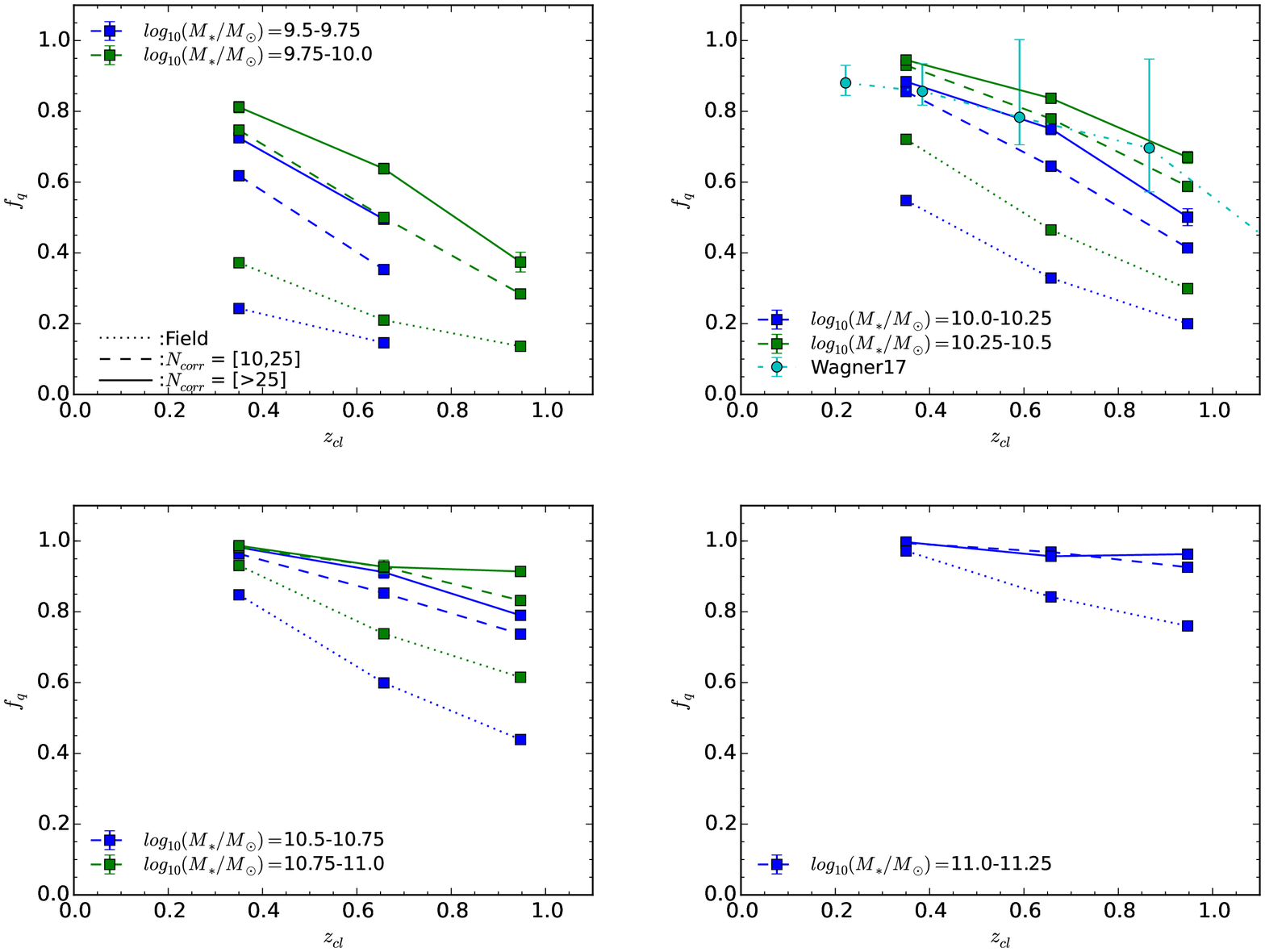}
\caption{The redshift evolution of the quiescent fractions ($f_q$s) for field (dotted), group (dashed), cluster (solid) galaxies, respectively.  Similar to the upper left panel in Figure~\ref{f4}, the data points in the high redshift bin for the mass bin 9.25-9.5 are removed due to the mass incompleteness. The Butcher-Oemler effect \citep{but84} is clearly seen but with a mass dependence. The effect is stronger for less massive galaxies and less for massive galaxies. For comparisons, we also add the $f_q$ results from \citet{wag17} (cyan circles) with M$_*$  $>$ 10$^{10.1}$ M$_{\odot}$ in the upper right panel. It is seen that our result is in agreement with that from \citet{wag17}. }
\label{f5}
\end{figure*}

To further quantitatively understand our results, the main sequence at different redshift is compared with that from \citet{whi12} in Figure~\ref{f2}. The purple, green, and blue solid lines represent the median SFR in three redshift ranges from low to high redshift in our sample, respectively. The three dashed lines plot the fitting lines at median redshifts in our three bins, $z_{med} = $ 0.33, 0.68, and 0.93, using the best-fit parameters from \citet{whi12}. We adopt the mass conversion factors from \citet{spe14} for stellar mass assuming different initial mass function (IMF), i.e., M$_{*,K}$  = 1.06 M$_{•,C}$ = 0.62 M$_{*,S}$, where S, C, and K, are referred to Salpeter, Chabrier, and Kroupa IMFs. The dotted blue line gives the best-fit results from \citet{noe07} in 0.2 $<$ $z$ $<$ 0.7. The yellow diamonds with error-bars are the SFRs of SF cluster galaxies in redshift range between 0.15 and 0.8 from \citet{muz12}, and the black and red triangles represent the SFRs of SF cluster galaxies in 0.15 $<$ $z$ $<$ 0.8 and 0.8 $<$ $z$ $<$ 1.5, respectively. It is seen that the best-fit slopes and magnitudes of the SF main sequence in our sample in three redshift ranges agree well with those in \citet{whi12} and \citet{noe07}, and are in broad agreement with those from \citet{wag17}. Additionally, we quantify the main sequence by fitting the data with a formula of a power law, i.e.
\begin{equation}\label{equ1}
log_{10}(\rm{SFR}/M_{\odot}   yr^{-1}) = \alpha log_{10}(M_*/M_{\odot}) + \beta,
\end{equation}
where $\alpha$ and $\beta$ are the slope and normalization, respectively.  In Figure~\ref{f3}, our best-fit $\alpha$ and $\beta$ for SF field galaxies in three different redshift ranges are plotted with a selection of literature best-fit parameters from a sample of 25 studies in the redshift range of 0 $<$ $z$ $<$ 6, compiled and calibrated by \citet{spe14}, including \citet{elb07}, \citet{noe07}, \citet{dun09}, \citet{san09}, \citet{oli10}, \citet{kar11}, \citet{whi12}, \citet{zah12}, and \citet{sob14}. All the above published best-fit parameters are standardized to a Kroupa IMF in \citet{spe14}. Besides, we add the results from \citet{wag17} for SF cluster galaxies, and from \citet{jian17} for SF field galaxies.  Our best-fit parameters are in good agreement with those from other published works, indicating the robustness of our results.

\begin{figure*}
\includegraphics[scale=0.6]{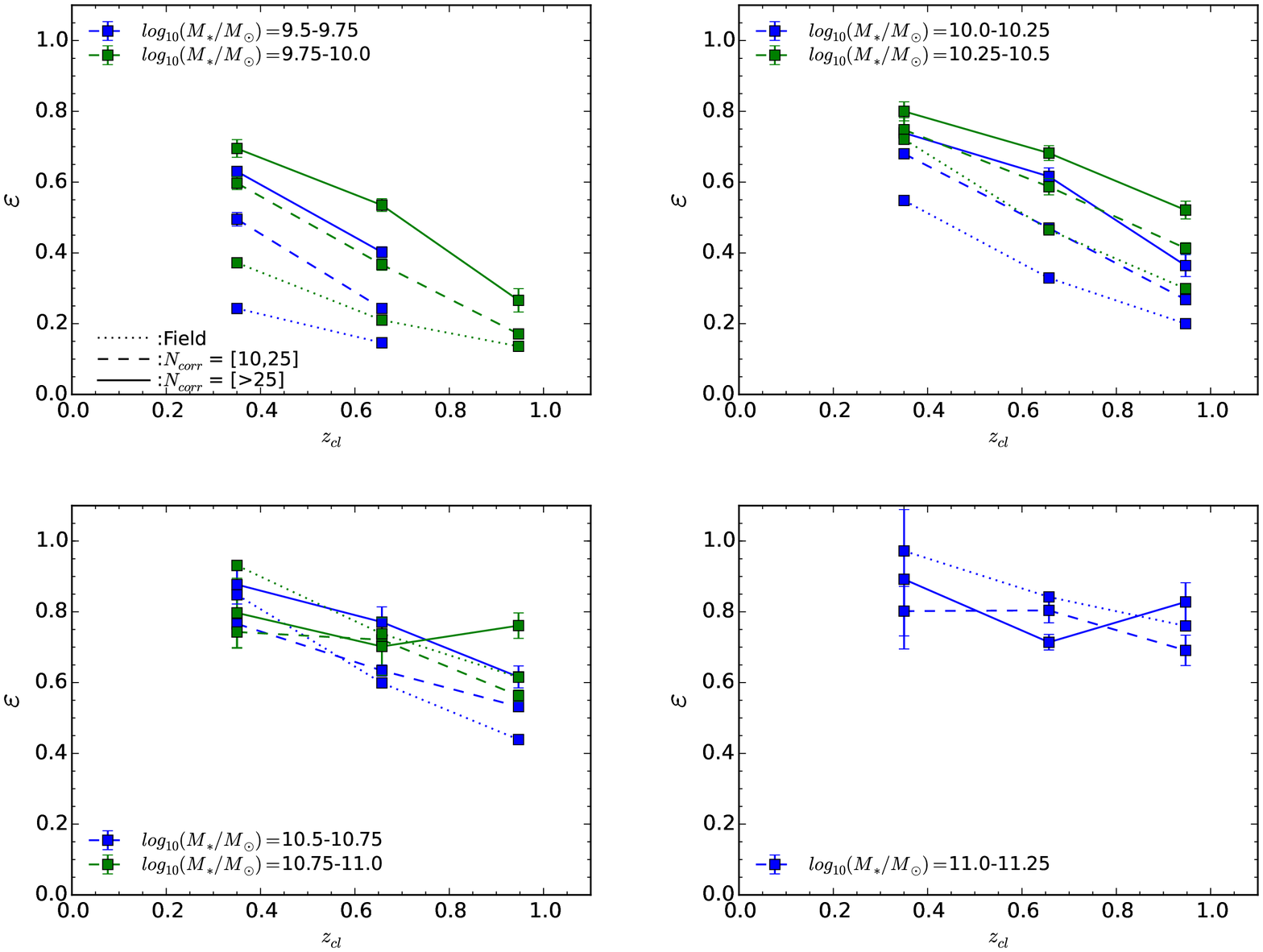}
\caption{Similar to Figure~\ref{f4} and \ref{f5}, the quenching efficiency $\varepsilon$ is plotted as a function of redshift for different mass ranges in four panels for field (dotted), group (dashed), and cluster (solid) galaxies, respectively. In general, the slope of the quenching efficiency decreases with the increasing masses, implying that the quenching effect is stronger for less massive galaxies.  }
\label{f6}
\end{figure*}

\subsection{Redshift Dependence}

\subsubsection{sSFR of SF galaxies, sSFR$_{\rm{SF}}$($z$)}
There have been many previous studies to discuss and compare the evolution of the main sequence in different environments \citep{koy13,lin14,erf16,wag17,jian17}. Taking the advantage of the large and deep HSC sample, we are able to explore the same subject with better statistics and low mass galaxies.  In Figure~\ref{f4}, the median sSFRs of the star-forming galaxies in field (dotted lines), group (dashed lines), and cluster (solid line) environments on the top, and the sSFR ratio of group to field galaxies (dashed lines) and that of cluster to field galaxies (solid lines) on the bottom are plotted as a function of redshift in different mass ranges separated in four panels. The yellow shaded region denotes the infrared main sequence as defined in \citet{elb11}, and is in the range of 13(13.8 Gyr - t )$^{-2.2}$ $<$  sSFR  $<$ 52(13.8 Gyr - t )$^{-2.2}$, where t is the look-back time. The red data circles mark the sSFR of the star-forming galaxies in the field from \citet{alb14} with stellar mass limit M $>$ 1.3 $\times$ 10$^{10}$ M$_{\odot}$. Four cyan circles with error-bars are from \citet{wag17} for SF cluster galaxies with mass larger than log$_{10}$(M$_*$/M$_{\odot}$) = 10.1. From Figure~\ref{f4}, our results show good consistency with that in \citet{elb11} and slightly shallower at low $z$ than that in \citet{alb14}.  It is seen that the decrease of the sSFR of the SF main sequence for field galaxies is $\sim$  0.45 dex from $z$ = 1.1 to 0.2 for all mass bins, possibly due to a global decline in the gas contents. At fixed masses, the decrease level is very similar to cluster and group galaxies between the same redshift range but a systematics is seen $<$ 0.4 dex for clusters and $<$ 0.2 dex for groups. That is, the decrease of sSFRs for the group to field galaxies and for the cluster to field ones is independent of redshift. However, the decrease appears to depend on stellar mass, being larger for less massive galaxies. Our results are in agreement with those in \citet{koy13} that the difference in the SFR-M$_*$ relation between cluster and field SF galaxies is $\sim$ 0.2 dex since $z$ $\sim$ 2.

\begin{figure*}
\includegraphics[scale=0.6]{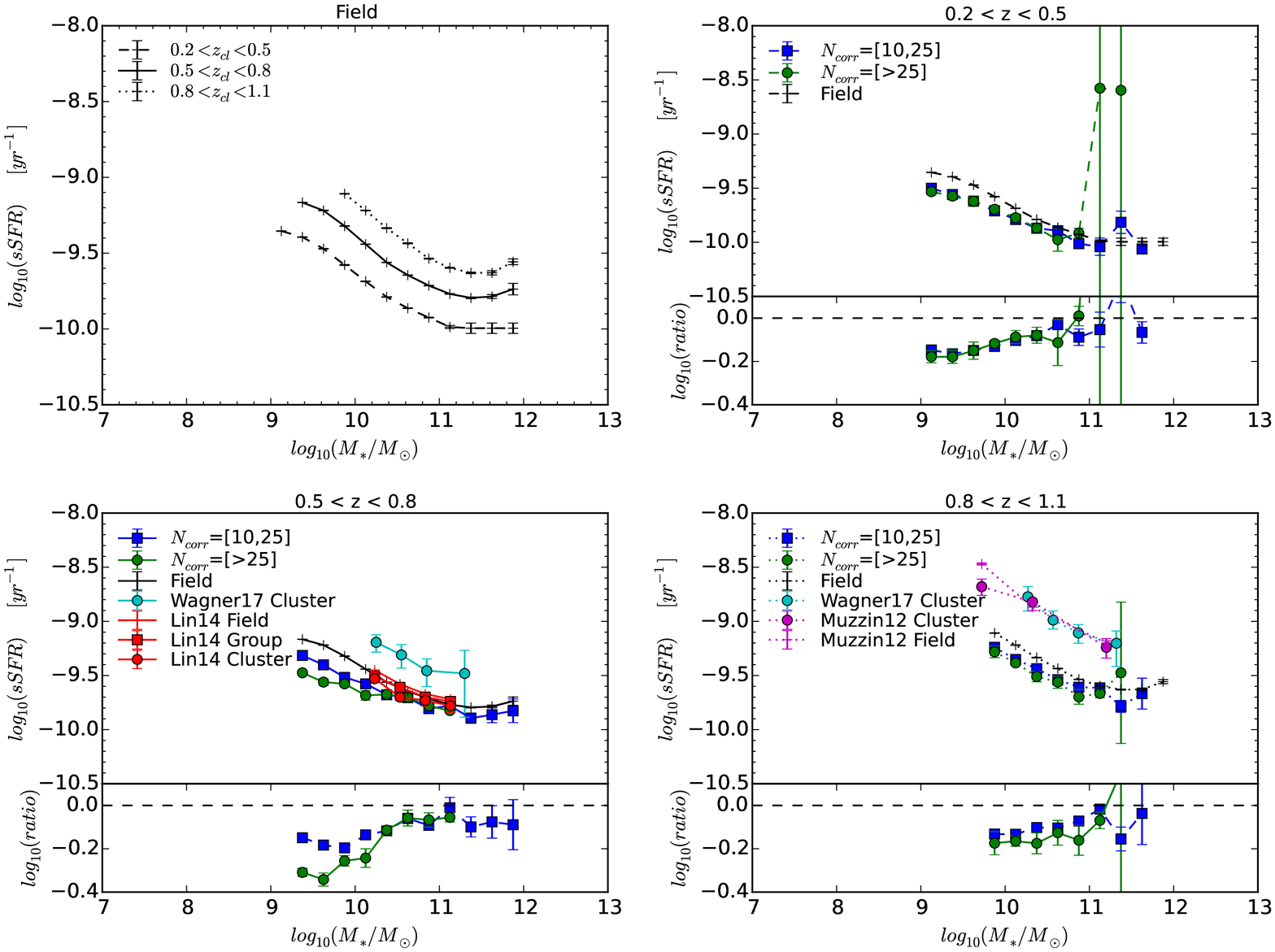}
\caption{Similar to Figure~\ref{f4}, but the sSFRs of SF galaxies are shown as a function of stellar mass. In the upper left panel, the sSFR of SF field galaxies in redshift range of 0.2 $<$ $z$ $<$ 0.5, 0.5 $<$ $z$ $<$ 0.8, and 0.8 $<$ $z$ $<$ 1.1 is represented by the dashed, solid, and dotted line, respectively. The other three panels plot the sSFRs of SF field (black pluses), group (blue squares), and cluster (green circles) galaxies on the top, and the logarithmic ratio of sSFR for group to field galaxies (blue squares), and for cluster to field galaxies (green circles) on the bottom. For comparisons, we also include the sSFR results from \citet{lin14} in the redshift range of 0.5 $<$ $z$ $<$ 0.8 for SF field (red pluses), group (red squares), and cluster (red circles) galaxies. In addition, the sSFR results are added as well from \citet{muz12} for SF field (purple pluses) and cluster (purple circles) galaxies in redshift range of 0.85 $<$ $z$ $<$ 1.2, and from \citet{wag17} for SF cluster galaxies (cyan circles) in redshift range of 0.15 $<$ $z$ $<$ 0.8 (bottom left) and 0.8 $<$ $z$ $<$ 1.5 (bottom right). }
\label{f7}
\end{figure*}

\subsubsection{Quiescent Fraction, $f_q$($z$)}
Under the stellar mass control, the redshift dependence of the quiescent fractions ($f_q$ )for field (dotted lines), group (dashed lines), and cluster (solid lines) galaxies is explored in Figure~\ref{f5}.  For comparison, the results from \citet{wag17} (cyan circles) with M$_*$ $>$ 10$^{10.1}$ M$_{\odot}$ are also plotted in Figure~\ref{f5}. Globally, the $f_q$ increases with decreasing redshift, consistent with the Butcher-Oemler effect \citep{but84}. However, depending on the galaxy masses, their effect can be significantly different. For high-mass galaxies, the Butcher-Oemler effect is weak and nearly negligible. On the other hand, the effect is stronger for low-mass ones, consistent with results in \citet{li12}.  For low-mass galaxies, the increase of $f_q$ is about 3 times larger from high $z$ $\sim$ 0.95 to low $z$ $\sim$ 0.35. In addition, the cluster downsizing effect that clusters with larger halo mass show greater quiescent fraction is also evident in our sample.   Our redshift evolution trend is similar to that in \citet{wag17},   and the $f_q$ derived for our sample in log$_{10}$(M$_*$/M$_{odot}$) = 10.2-10.5 (the blue solid line in upper right panel) appears to also be consistent with those in \citet{wag17}.    

\subsubsection{Quenching Efficiency, $\varepsilon$($z$)}
We also compute the environmental quenching efficiency, $\varepsilon^{\textrm{envi}}$ = ($f^{\textrm{cluster}}_{q} - f^{\textrm{field}}_{q}$)/(1 - $f^{\textrm{field}}_{q}$), similar to the definition in \citep{pen10}, as a function of redshift in different mass bins to quantify the excess of quenching due to pure environmental effects in Figure~\ref{f6}.  In general, the quenching efficiency increases with the decreasing redshift, and also increases with the increasing mass at fixed redshift. However, the slope of the quenching efficiency decreases with the increasing masses, implying that the quenching effect is stronger for less massive galaxies. The redshift dependence becomes weaker for the high-mass galaxies, likely due to being red and dead already for most of the massive galaxies at high redshifts.  In addition, it is seen that the environmental quenching in clusters is larger than that in groups, indicating that the hosting environment has a significant effect on the star formation quenching.

\begin{figure*}
\includegraphics[scale=0.6]{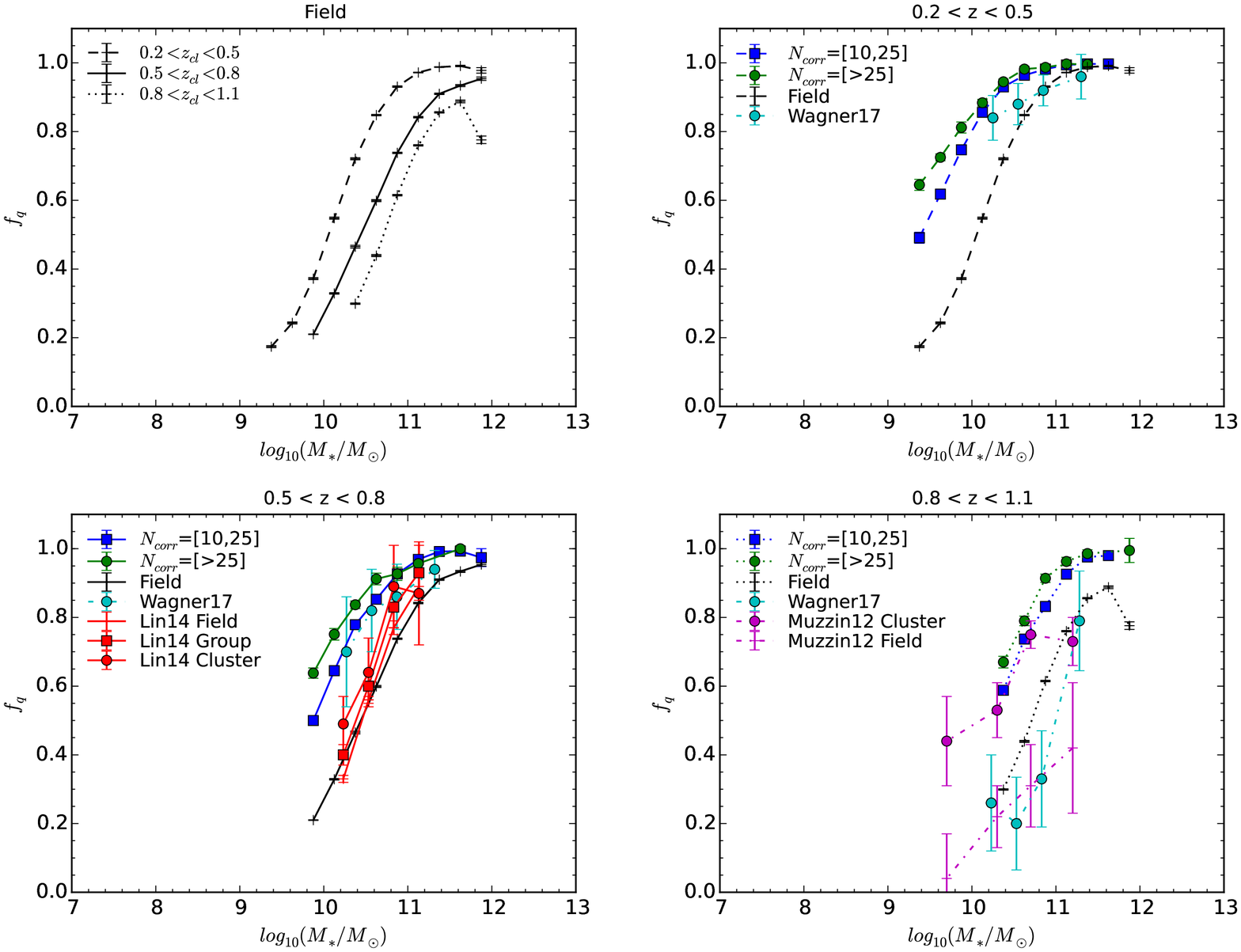}
\caption{The quiescent fraction $f_q$s as a function of stellar mass. In upper left panel, the f$_q$s are plotted for field galaxies in three redshift bins while the other three panels show the f$_q$s of field (black pluses), group (blue squares), and cluster (green circles) galaxies in redshift range of 0.2 $<$ $z$ $<$ 0.5 (upper right), 0.5 $<$ $z$ $<$ 0.8 (bottom left), and 0.8 $<$ $z$ $<$ 1.1(bottom right). The red pluses, squares, and circles denote the $f_q$s for field, group, and cluster galaxies, respectively, from \citet{lin14} in redshift range of 0.5 $<$ $z$ $<$ 0.8. The cyan circles are data points from \citet{wag17} for cluster galaxies in redshift range of 0.15 $<$ $z$ $<$ 0.41 (upper right), 0.41 $<$ $z$ $<$ 0.8 (bottom left) and 0.8 $<$ $z$ $<$ 1.5 (bottom right). The purple pluses and circles represent the $f_q$s for field and cluster galaxies, respectively, from \citet{muz12} in redshift range of 0.85 $<$ $z$ $<$ 1.2.}
\label{f8}
\end{figure*}

\subsection{Stellar Mass Dependence}
\subsubsection{sSRF of SF galaxies, sSFR$_{\rm{SF}}$(M$_*$)} \label{ssfr_ms}
In Figure~\ref{f7}, the sSFRs of SF galaxies are plotted as a function of the stellar mass M$_{*}$ in different hosting environments and redshift ranges. In the upper left panel, the median sSFR of SF field galaxies is shown in redshift range of 0.2 $<$ $z$ $<$ 0.5 (dashed line), 0.5 $<$ $z$ $<$ 0.8 (solid one), and 0.8 $<$ $z$ $<$ 1.1 (dotted line) for comparisons, respectively. For the other three panels on top-right, bottom-left, and bottom-right, the sSFRs of SF field (black pluses), group (blue squares), cluster (green circles) galaxies are plotted on the top while the logarithmic ratio of sSFR for SF group to field galaxies (blue squares) and that for SF cluster to field ones (green circles) are displayed on the bottom. Additionally, the red pluses, squares, and circles denote the sSFR of SF field, group, and cluster galaxies in 0.5 $<$ $z$ $<$ 0.8 from \citet{lin14}, respectively. The cyan circles are the sSFRs of SF cluster galaxies in 0.15 $<$ $z$ $<$ 0.8 and in 0.8 $<$ $z$ $<$ 1.5 from \citet{wag17}, and the purple pluses and circles are those of SF field and cluster galaxies in 0.85 $<$ $z$ $<$ 1.2 from \citet{muz12}.   

In general, the median sSFR of group or cluster galaxies decreases from that of field ones, implying a slow quenching effect acting in dense environments. We found that there is a weak dependence on the stellar mass of the sSFR deficit relative to the field, which is $<$ 0.2 dex in groups and  $<$ 0.4 dex in clusters. The larger deficit seen in cluster galaxies as opposed to group galaxies suggests a stronger environmental quenching effect in clusters than in groups. It is worth noting that the decreases are slightly larger from what was found in \citet{lin14}, who found no significant sSFR reduction of SF galaxies in the groups and $\sim$ 17 $\%$ (0.23 dex) sSFR decrease for the SF galaxies in the clusters relative to the field galaxies. However, we note that the halo mass binning is 10$^{13.6-14.2}$ M$_{\odot}$ for groups and 10$^{ > 14.2}$ M$_{\odot}$ for clusters in this work while it is 10$^{13.4-14.0}$ M$_{\odot}$ for groups and 10$^{> 14.0}$ M$_{\odot}$ for clusters in \citet{lin14}. The larger sSFR offset seen in this work may be attributed to the greater masses of the groups and clusters probed in our analysis, since larger sSFR reduction is expected in more massive clusters. In addition, less massive groups normally suffer from the more serious foreground and background contaminations and the real group signals can be easily smeared out, leading to the less sSFR reduction. In the high redshift (bottom-right) panel, although there seems to have a systematic between our results and the results from \citet{muz12} and \citet{wag17} for the absolute sSFR values, the decreasing trend of the sSFR with the increasing mass in our sample agrees well with them. Besides, it is noticed that our results for the sSFR reductions are in broad agreement with those from \citet{muz12}. 

\begin{figure*}
\includegraphics[scale=0.6]{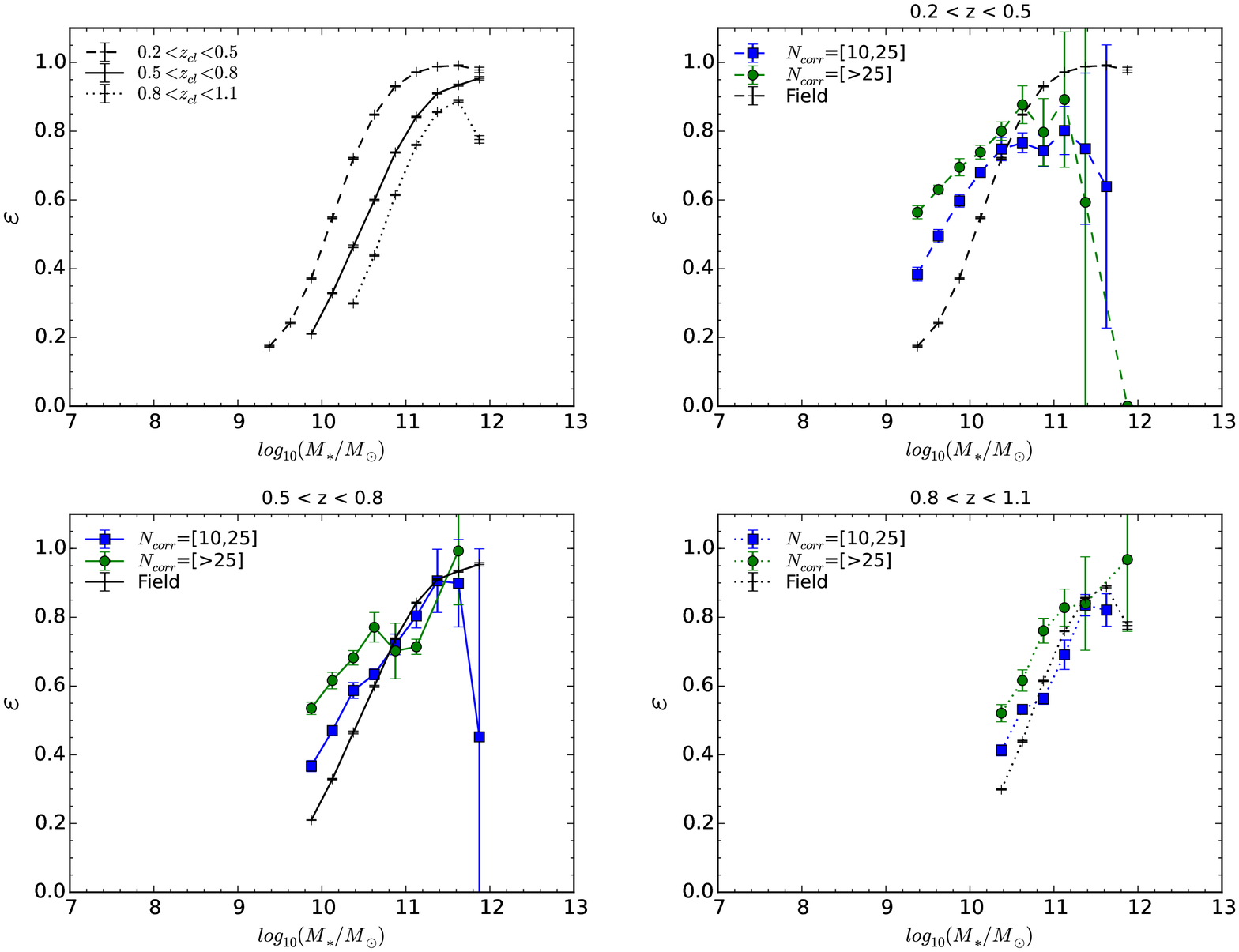}
\caption{The quenching efficiency $\varepsilon$s as a function of stellar mass. Similar to Figure~\ref{f7} and \ref{f8}, the $\varepsilon$s of field galaxies in three redshift bins are shown in the upper left panel while the other three panels plot the $\varepsilon$s of field (black pluses), group (blue squares), and cluster galaxies (green circles) in three redshift ranges, respectively. It is clearly seen that the transition masses from the dominance of the environment quenching to that of the mass quenching are log$_{10}$(M$_*$/M$_{\odot}$) $\sim$ 10.4 in groups and $\sim$ 10.6 in clusters.}
\label{f9}
\end{figure*}

\subsubsection{Quiescent Fraction, $f_q$(M$_*$)}
The quiescent fraction ($f_q$) is investigated as a function of $M_*$ in Figure~\ref{f8}. Similar to Figure~\ref{f7}, the upper-left panel shows the $f_q$ for field galaxies in three redshift ranges, and the other three panels give the comparisons of $f_q$ for the field, group, and cluster galaxies in one redshift range per one panel. Results of $f_q$ are included for comparisons from \citet{muz12} for field and cluster galaxies in redshift range of 0.85 $<$ $z$ $<$ 1.2, from \citet{lin14} for field, group, and cluster galaxies in 0.5 $<$ $z$ $<$ 0.8, and from \citet{wag17} for cluster galaxies in 0.15 $<$ $z$ $<$ 0.41, 0.41 $<$ $z$ $<$ 0.8, and 0.8 $<$ $z$ $<$ 1.5. It is seen that the quiescent fraction depends strongly on mass and less strongly on redshift and the hosting environment. At high redshifts, the $f_q$ difference at fixed mass between group and cluster galaxies is small and gradually becomes larger with the decreasing redshift, indicating that the environmental quenching effect has a redshift dependence.  In addition, the typical “downsizing” effect in which less massive galaxies are more star-forming continues down to $\sim$ $10^{9}$ M$_{\odot}$. Moreover, at low mass, it is found that the red sequence is still visible in groups or clusters with respective to the field, implying that the environmental quenching can still effectively stop star-formation for low mass galaxies. It is noticed that the $f_q$ in this work is higher compared to those from \citet{lin14} and \citet{wag17}. Besides the greater cluster masses used in this work (see the discussion in the previous section \ref{ssfr_ms}), it is also likely that the CAMIRA cluster finding is based on the red sequence method and preferentially select clusters with high $f_q$. In the high redshift range,  our $f_q$s broadly agree with those from \citet{muz12}. However, lower $f_q$s are found in \citet{wag17}, likely due to the fact that their sample includes galaxies up to $z$ $\sim$1.5.

\subsubsection{Quenching Efficiency, $\varepsilon$(M$_*$)}
The environmental quenching efficiency is schemed as a function of stellar mass in Figure~\ref{f9} in order to better quantify the excess of quenching due to pure environmental effects. The panel arrangement is similar to that in Figure~\ref{f7} and \ref{f8}. In the upper-left panel, the quenching efficiency from field galaxies, or the mass quenching efficiency assuming that field galaxies are purely quenched by their masses, are displayed in three redshift ranges. The other three panels present the comparisons among the field, group, and cluster galaxies per redshift bin. It is clearly seen that both the mass and environmental quenching increases with increasing stellar mass. However, high mass galaxies are dominated by the mass quenching while low mass galaxies are quenched primarily due to the environmental quenching. The transition mass where the dominance of the mass and environmental quenching switches is recognized at log$_{10}$(M$_*$/M$_{\odot}$) $\sim$ 10.4 in groups and 10.6 in clusters in the lowest redshift range of $0.2 < z < 0.5$, in good agreement with the results from \citet{lin14}.  This implies that the environmental quenching in clusters is more prominent than that in groups since it can quench more massive galaxies than in groups.

\section{Conclusion and Discussion}
In this work, we probe galaxies down to $i \sim$ 26, two magnitudes deeper than that in our previous works \citep{lin14,jian17}, and extend our results to redshift 1.1, enabling us to study the cosmic evolution of the star formation activity. Due to the exceptional deep depth of our sample, the stellar mass completeness limit can be pushed to log$_{10}$(M$_*$/M$_{\odot}$) = 9.8 and 10.3 at $z$ $\sim$ 1.1 for star-forming and quiescent galaxies, respectively. That is, we are able to explore the quenching status of lower-mass star-forming galaxies which have not explored in our previous works. Additionally, the HSC Wide 16A data are collected from a survey area $\sim$ 174 square degree to offer good statistics even for galaxies at the high-mass end. In other words, we can constrain the results on high mass galaxies more robustly as opposed to previous studies. The main results can be summarized below:

1. The sSFR of star-forming galaxies decreases with cosmic time, regardless the environments. The sSFR is lower in groups and clusters than in the field by $\sim$0.1 dex and $\sim$0.2--0.3 dex, respectively. The drop of sSFR in group/cluster environments is insensitive to the redshift and is more significant for low mass galaxies.

2. The quiescent fraction is a strong function of stellar mass in all environments, being greater for higher mass galaxies. The Butcher-Oemler effects are seen in all environments and are more prominent for low mass galaxies. 

3. Both the environment and stellar mass quenching efficiencies increase with the stellar mass. However, the environment quenching dominates over the stellar mass quenching for low mass galaxies. 

At a fixed stellar mass, we find that the decrease in the sSFR of star-forming group or cluster galaxies to field ones is 0.1-0.3 dex and is more prominent in cluster environments. This implies that galaxies in clusters likely experience a long timescale (or slow) quenching effect that gradually reduces the SFR of star-forming galaxies. Although the gas contents and hence the star formation rate of galaxies vary with redshifts, we find that the offsets in the sSFR in groups/clusters, as opposed to the field, are comparable at different redshifts, in agreement with the results found by \citet{koy13}. This implies that the slow quenching mechanism acting on groups/clusters is likely universal in the redshift range of $0.2 < z < 1.1$.        

In addition, it is found in this work that the transition masses from environment quenching to the mass quenching are 10$^{10.4}$ M$_{\odot}$ and 10$^{10.6}$ M$_{\odot}$  in groups and clusters, respectively, in the lowest redshift bin of $0.2 < z < 0.5$, in good agreement with the results from \citet{lin14}. The greater transition mass in clusters suggests that the environment effects are more important in clusters than in groups and have an effect even for massive galaxies. It is also noticed that the red fraction for the most massive galaxies is comparable between the field and group/cluster environments. This suggests that the star formation of those massive galaxies beyond the transition mass are likely already stopped and being red and dead before they enter the cluster-like environments. Conversely, the environmental quenching dominates the mass quenching in low mass galaxies below the transition mass in groups or clusters down to the mass completeness limit of 10$^{8.6}$ M$_{\odot}$ at $z$ $\sim$ 0.2, and the environmental quenching effect in clusters is stronger than that in groups. At this low mass regime, the red sequence is still visible in groups or clusters with respective to the field, suggesting that the environmental quenching can still effectively stop star-formation for low mass galaxies.     

\citet{koy17} recently present their results discussing the environmental dependence of color, stellar mass, and sSFR of H$_{\alpha}$-selected galaxies in twin clusters in the DEEP2-3 field at $z$ = 0.4. One import finding in their work is that H-selected galaxies reveal color-density or color-radius correlations, but their stellar masses or sSFRs are independent of environments. The conclusion appears to be inconsistent with our result that there is a systematical reduction of sSFR for SF galaxies in groups or clusters by 0.1 - 0.3 dex with respect to those in the field. As discussed in \citet{koy17}, there are various factors that may contribute to this discrepancy. It is likely due to different definitions for SF galaxies in our and their sample. Our SF galaxies are determined by their sSFRs while SF galaxies are detected as H$_{\alpha}$ galaxies in \citet{koy17}. In addition, it is indicated that the treatment of dust extinction or NII line contamination could easily change the results by $>$10-20$\%$ level, which can significantly ease the inconsistency between us. Finally, it is also possible that the sample size of H$_{\alpha}$ emitters in \citet{koy17} is much smaller than that of our sample, which covers whole HSC S16A Wide survey area, and poor statistics (particularly in a high-density environment) may lead to the disagreement in conclusions.

\begin{ack}

This work is supported by the Ministry of Science \& Technology of Taiwan under the grant MOST 103-2112-M-001-031-MY3 and MO is supported in part by World Premier International Research Center Initiative (WPI Initiative), MEXT, Japan, and JSPS KAKENHI Grant Number 26800093 and 15H05892. The Hyper Suprime-Cam (HSC) collaboration includes the astronomical communities of Japan and Taiwan, and Princeton University.  The HSC instrumentation and software were developed by the National Astronomical Observatory of Japan (NAOJ), the Kavli Institute for the Physics and Mathematics of the Universe (Kavli IPMU), the University of Tokyo, the High Energy Accelerator Research Organization (KEK), the Academia Sinica Institute for Astronomy and Astrophysics in Taiwan (ASIAA), and Princeton University.  Funding was contributed by the FIRST program from Japanese Cabinet Office, the Ministry of Education, Culture, Sports, Science and Technology (MEXT), the Japan Society for the Promotion of Science (JSPS),  Japan Science and Technology Agency  (JST),  the Toray Science  Foundation, NAOJ, Kavli IPMU, KEK, ASIAA,  and Princeton University.
\end{ack}

\end{document}